\documentclass[
12pt,tightenlines, amsmath, amssymb, nofootinbib, prd,
superscriptaddress, showpacs, preprintnumbers]{revtex4}
\newcommand{\beq}{\begin{equation}}
\newcommand{\eeq}{\end{equation}}
\newcommand{\ba}{\begin{array}{ccc}}
\newcommand{\ea}{\end{array}}
\newcommand{\nn}{\nonumber}
 \renewcommand{\d}{\partial}
\def\beqn{\begin{eqnarray}}
\def\eeqn{\end{eqnarray}}

\def\Tr{ {\rm Tr} }
\def\<{\langle}
\def\>{\rangle}
\def\pp{{\langle\bar{\psi}\psi\rangle}_0}
\def\cond{ \frac{b g^2}{32 \pi^2 } G_{\mu \nu}^a G^{\mu \nu a}}
\def\GDG{\frac{g^2 G \tilde{G}}{32 \pi^2}}
\usepackage{graphicx}
\usepackage{amsmath}
\usepackage{amssymb}
\usepackage{psfrag}
\usepackage{epsfig}
\begin{document}
\title{\Large{$\theta$-dependence of $QCD$ at Finite Isospin Density}}
 \affiliation{Department of Physics and
Astronomy, University of British Columbia, Vancouver, BC, Canada,
V6T 1Z1}
 \author{ Max~A.~Metlitski}
 \email{mmetlits@phas.ubc.ca}
 \affiliation{Department of Physics and
Astronomy, University of British Columbia, Vancouver, BC, Canada,
V6T 1Z1}
 \author{ Ariel~R.~Zhitnitsky}
 \email{arz@phas.ubc.ca}
\affiliation{Department of Physics and Astronomy, University of
British Columbia, Vancouver, BC, Canada, V6T 1Z1}

\date{\today }

\vfill
\begin{abstract}
We probe the $\theta$-dependence of $QCD$ at finite isospin
chemical potential $\mu_I$ using the effective chiral Lagrangian
approach. The phase diagram in the $\theta$, $\mu_I$ plane is
constructed and described in detail in terms of chiral and pion
condensates. The physics at $\theta \sim \pi$ is investigated in
both the normal and superfluid phase. Finally, the behaviour of
the gluon condensate at finite $\mu_I$ is computed.
\end{abstract}

\vfill

\maketitle

\section{Introduction}
The $\theta$-parameter of gauge theories has long attracted
attention as it is a probe of the topological properties of the
theory. In almost every context, from pure Yang-Mills theories to
$QCD$ the $\theta$ dependence of the theory is highly non-trivial
and, frequently, non-analytic. In particular, in $QCD$ with two
flavours and equal non-zero quark masses it is believed that the
so called Dashen's
phenomenon\cite{Dashen,Witten:1980sp,DiVecchia:1980ve,Creutz,Smilga,Tytgat}
- a first order phase transition, characterized by spontaneous
breaking of $CP$, occurs at $\theta = \pi$.

In this paper, we investigate the influence of finite isospin
chemical potential $\mu_I$ on the $\theta$ dependence of two
flavor $QCD$. Besides pure academic interest, the main physical
motivation for such a study is the attempt to understand the
cosmological phase transition when $\theta$, being non-zero and
large at the very beginning of the phase transition, slowly
relaxes to zero, as the axion resolution of the strong CP problem
suggests. Of course, in real world, we are mostly interested in
the effects of $\theta$ on matter at finite baryon, rather than
isospin, density. Indeed, if isospin asymmetric matter presently
exists in nature (say in neutron stars), it is accompanied by a
large baryon density. However, analytical control over $QCD$ is
absent at moderate baryon density and appears only at
asymptotically large baryon chemical potential, where one expects
the color-superconducting state to be realized\cite{ARF}.
Nevertheless, one may resort to $QCD$-like theories, such as $N_c
= 2$ $QCD$ at finite baryon
density\cite{kogut1,kogut2,Splittorff:2000mm} and $N_c = 3$ $QCD$
at finite isospin density\cite{Son:2000xc,Son:2000by}, where
analytical control is present, to gain some insight into real
dense $QCD$.

Due to the axial anomaly, the $\theta$ parameter of $QCD$ is
intimately tied to the quark mass matrix and may be incorporated
into the effective chiral
Lagrangian\cite{Witten:1980sp,DiVecchia:1980ve}. There also exists
a well-known procedure for including the effects of finite $\mu_I$
into the $QCD$ chiral Lagrangian\cite{Son:2000xc,Son:2000by}. We,
thus, expect that we may adequately describe $QCD$ at finite
isospin density and $\theta \neq 0$ in the effective Lagrangian
approach, as long as $\mu_I$ is much smaller than the mass of the
lightest non-Goldstone boson (in $QCD$, the mass of the $\rho$
meson, $m_\rho$).

Using the above approach, we obtain a wide range of information
about the phase diagram of two flavor $QCD$ in the $\mu_I$,
$\theta$ plane. We show that the transition to the superfluid,
isospin breaking, phase occurs at $\mu_I$ equal to the $\theta$
dependent pion mass, $m_\pi(\theta)$. This implies that for fixed
$\mu_I$ of order of the pion mass, the $\theta$ dependence of the
theory becomes non-analytic. Two second order phase transitions,
accompanied by a jump in the topological susceptibility, occur as
$\theta$ relaxes from $2 \pi$ to $0$.

We compute the $\theta$ dependence of chiral and pion condensates,
as well as $\langle i G \tilde{G}\rangle$ and the topological
susceptibility, in normal and superfluid phases. We find that the
$\theta$ dependence in the superfluid phase near $\theta = \pi$ is
much smoother than in the normal phase. In particular, for $m_u =
m_d$, we show that the first order phase transition across $\theta
= \pi$ present in the normal phase, disappears in the superfluid
phase.

Finally, we discuss a few $\theta$ unrelated issues. Most
importantly, we compute the dependence of the gluon condensate
$\langle \cond \rangle$ on the isospin chemical potential in the
superfluid phase. The gluon condensate decreases with density near
the normal to superfluid phase transition, but,
counter-intuitively, increases for $m_\pi \ll \mu_I \ll m_{\rho}$.
We also  evaluate a novel vacuum expectation value, which appears
in the superfluid phase:
  $\langle i \bar{u} \gamma_0 \gamma_5 d\rangle$. This density, being nonzero even at $\theta=0$, nonetheless
 has never been discussed in the literature previously.
 This density, itself, breaks the isospin symmetry, and so may be
considered as an additional order parameter.

We note that the above agenda has also recently been implemented
to study the properties of $N_c = N_f = 2$ $QCD$ in the presence
of non-zero $\theta$ at finite baryon and isospin density. Most of
the results of the present study are in direct correspondence with
the work\cite{MZ}. This is a consequence of the fact that the
chiral Lagrangians describing $N_c = 3, N_f = 2$ $QCD$ and the
pion sector of the $N_c = 2, N_f = 2$ $QCD$ are identical. Besides
adapting the work\cite{MZ} to the $N_c = 3$ context, we presently
discuss in some detail the theoretically interesting case of
exactly degenerate quark masses, which was not analyzed in
\cite{MZ}.


We hope that the results of this study would be of interest for
lattice simulations. Indeed, the determinant of the Dirac operator
is real and positive in $QCD$ at non-zero isospin chemical
potential and $\theta = 0$. The determinant remains real at $\mu_I
\neq 0$, $\theta = \pi$. Thus, we hope that the $\mu_I$ dependence
of the gluon condensate and the topological susceptibility at
$\theta = 0$, can be explicitly checked on the lattice. This is a
unique chance to study the gluon degrees of freedom and their
dependence on light quark masses. The corresponding study might be
important for the extrapolation procedure which has to be used in
order to achieve the chiral limit. Moreover, we hope that the
disappearance of Dashen's phenomenon at $\theta = \pi$ in the
superfluid phase can also be confirmed by lattice simulations.

\section{The Chiral Lagrangian}
The low energy dynamics of $N_f = 2$ $QCD$  are governed by the
chiral Lagrangian for the pion field $U \in SU(2)$. A well known
procedure exists to incorporate into this Lagrangian the effects
of a finite $\theta$
parameter\cite{Witten:1980sp,DiVecchia:1980ve}. A method for
introducing a finite isospin chemical potential $\mu_I$ is also
well-developed\cite{Son:2000xc,Son:2000by}. To lowest order in
quark mass and derivatives, the chiral Lagrangian reads, \beq
\label{Lch} {\cal L} = \frac{1}{4} f^2_{\pi} \Tr(\nabla_{\mu}U
\nabla_{\mu}U^{\dagger}) - \Sigma Re\Tr(M U) \eeq where the flavor
covariant derivatives are defined as, \beqn \nabla_0
U &=& \d_0 U - \frac{1}{2}\mu_I [\tau^3, U], \quad \nabla_i U = \d_i U\\
\nabla_0 U^{\dagger} &=& \d_0 U + \frac{1}{2}\mu_I [U^{\dagger},
\tau^3], \quad \nabla_i U^{\dagger} = \d_i U^{\dagger}\eeqn We
work in Euclidean space. Here the $\theta$ parameter of $QCD$ has
been incorporated directly into the quark mass matrix, \beqn M =
e^{-i\theta/N_f}\left(\begin{array}{cc} m_u & 0\\0 & m_d
\end{array}\right)\eeqn We keep $m_u\neq m_d$ on purpose: as
is known $m_u=m_d$ is a very singular
 limit when one discusses $\theta$ dependence, see below.
The coefficient $\Sigma$ is determined by the chiral condensate in
the limit $m\rightarrow 0^+$, $\theta = 0$, $\mu_I = 0$, \beq
\Sigma = -\frac{{\langle\bar{\psi}\psi\rangle}_0}{2 N_f}\eeq as
will be confirmed below. In our notations the chiral condensate
includes the sum over all flavors, $\langle \bar{\psi}\psi\rangle
= \sum_f{\langle\bar{\psi}_f\psi_f\rangle}$.

Due to pseudo-reality of $SU(N_f = 2)$, one may, to this order in
chiral perturbation theory, incorporate all effects of $\theta$,
$m_u$, $m_d$ into a common real quark mass via a redefinition,
\beqn U &=& L \tilde{U} R^{\dagger}, \quad L = R^{\dagger} = e^{i
\alpha
\tau^3/2}\\
\cos{\alpha} &=& \frac{(m_u + m_d)
\cos(\theta/2)}{\sqrt{(m_u+m_d)^2 \cos^2(\theta/2) + (m_u-m_d)^2
\sin^2(\theta/2)}}\nn\\
\sin\alpha &=&  \frac{(m_u - m_d)
\sin(\theta/2)}{\sqrt{(m_u+m_d)^2 \cos^2(\theta/2) + (m_u-m_d)^2
\sin^2(\theta/2)}}\label{U0}\eeqn Our parameter $\alpha$ is
related to the commonly used Witten's variables $\phi_u,
\phi_d$\cite{Witten:1980sp}, via, \beqn \phi_u = \theta/2 -
\alpha, \quad \phi_d = \theta/2 + \alpha\\ \phi_u + \phi_d =
\theta, \quad m_u \sin \phi_u = m_d \sin \phi_d\eeqn After such a
transformation, the Lagrangian (\ref{Lch}) takes the form, \beq
\label{Ls} {\cal L} = \frac{1}{4} f^2_{\pi}
\Tr(\nabla_{\mu}\tilde{U} \nabla_{\mu}\tilde{U}^{\dagger}) -
m(\theta)\Sigma \, Re\Tr(\tilde{U})\eeq with, \beq \label{m}
m(\theta) = \frac{1}{2}\left({(m_u+m_d)^2\cos^2(\theta/2) +
(m_u-m_d)^2\sin^2(\theta/2)}\right)^{\frac12}\eeq

\section{Phase Diagram}
Our next step is to find the classical minimum of the effective
Lagrangian (\ref{Ls}) to determine the phase diagram. First let's
study the theory at zero chemical potential and fixed $\theta$.
The classical minimum, is then given by, $\tilde{U} = 1$, and the
lowest lying excitations correspond to a triplet of pions, with
$\theta$ dependent mass, \beq \label{mpi} m^2_{\pi}(\theta) =
\frac{m(\theta)|\pp|}{f^2_{\pi}}\eeq The pion mass $m_{\pi}$
acquires a dependence on $\theta$ through the effective quark mass
parameter $m(\theta)$ (\ref{m}). 
As we shall see, the whole phase diagram turns out to be
determined by the parameter $m_{\pi}(\theta)$. We note that
$m_{\pi}(\theta)$ reaches its maximum at $\theta = 0$ and minimum
at $\theta = \pi$. Moreover, for $m_u = m_d$, $\theta = \pi$,
$m_\pi$ vanishes to first order in $m_q$.

Now let's turn on finite $\mu_I$. At fixed $\theta$, the phase
diagram contains two phases: normal and superfluid. The transition
from the normal phase to the superfluid phase occurs at the
critical chemical potential $\mu_I = m_\pi(\theta)$. In the normal
phase, $\langle\tilde{U}\rangle  = 1$. In the superfluid phase,
the $U(1)_I$ symmetry is spontaneously broken and, \beqn
\label{min} \langle\tilde{U}\rangle = \lambda(\theta) + i
\sqrt{1-\lambda(\theta)^2} (\tau^1 \cos \phi + \tau^2
\sin\phi)\eeqn where the variable $\phi$ labels the $U(1)_I$
degeneracy of the vacuum, and we have introduced the parameter
$\lambda$ to describe both the normal and superfluid phase, \beqn
\lambda(\theta) = \left\{
\begin{array}{cl}
1 &\quad \textrm{normal phase}\\
\frac{m^2_\pi(\theta)}{\mu^2_I} &\quad \textrm{superfluid phase.}
\end{array} \right.
\eeqn As expected, at $\theta=0$ we reproduce the known results
\cite{Son:2000xc,Son:2000by}. At $\theta \neq 0$ the phase diagram
looks the same as at $\theta = 0$, with the important replacement,
$m^2_\pi \rightarrow m^2_\pi(\theta)$. This is a very natural
conclusion. Indeed, at $\theta \neq 0$ pions still carry isospin
number. Hence, their energy is lowered at finite isospin chemical
potential. As soon as $\mu_I$ reaches the vacuum pion mass
$m_\pi(\theta)$, Bose-condensation occurs leading to spontaneous
breaking of $U(1)_I$ symmetry.

Quantitatively, the $\theta$ dependence of the Goldstone mass
$m_\pi(\theta)$ implies that the transition to superfluid phase is
shifted to a smaller chemical potential $\mu_I$, compared to
$\theta = 0$. In the limiting case, when $m_u = m_d$ and $\theta =
\pi$, the transition occurs in the vicinity of $\mu = 0$ (see
Section IV for a more precise discussion). For physical values,
$m_d = 7 MeV$, $m_u = 4 MeV$, the transition at $\theta = \pi$
occurs at $\mu = \left(\frac{m_d - m_u}{m_d +
m_u}\right)^{\frac12}m_\pi(0) \sim 70 MeV$.

We now wish to describe the phase diagram in terms of different
condensates and densities. This can be achieved by the standard
procedure of introducing sources into the chiral Lagrangian. We
find that chiral condensates depend on $\mu_I$, $\theta$ in the
following way, \beqn \label{cond1}\langle \bar{u} u \rangle &=&
\frac{1}{2} \pp \lambda(\theta)
\cos(\frac{\theta}{2}-\alpha),\quad \langle\bar{d} d \rangle =
\frac{1}{2} \pp \lambda(\theta) \cos(\frac{\theta}{2}+\alpha)\\
i\langle\bar{u} \gamma_5 u \rangle &=& -\frac{1}{2} \pp
\lambda(\theta) \sin(\frac{\theta}{2}-\alpha),\quad
i\langle\bar{d} \gamma_5 d \rangle = -\frac{1}{2} \pp
\lambda(\theta) \sin(\frac{\theta}{2}+\alpha)\nn\eeqn while the
pion condensate, which exists only in the superfluid phase and
spontaneously breaks the $U(1)_I$ symmetry, takes the form, \beqn
\label{cond2}i\langle\bar{u}\gamma_5 d \rangle = \frac{1}{2}\pp
\sqrt{1-\lambda^2(\theta)}\cos(\frac{\theta}{2}),\quad \langle
\bar{u}d\rangle = \frac{1}{2}\pp
\sqrt{1-\lambda^2(\theta)}\sin(\frac{\theta}{2})\nn\eeqn 
Notice that once $\theta\neq 0$, the $P$ odd condensate
$i\langle\bar{q} \gamma_5 q \rangle$ appears in addition to the
usual $P$ even condensate $\langle\bar{q}{q}\rangle$. Similarly,
in the superfluid phase at $\theta \neq 0$, the $P$ even
condensate $\langle \bar{u} d\rangle$ exists alongside the
ordinary $P$ odd pion condensate, $i\langle  \bar{u} \gamma_5
d\rangle$. This is a direct consequence of explicit parity
violation by the $\theta$ term.

We may also compute the following charge densities from our chiral
Lagrangian, \beqn n_I &=& \frac{1}{2} \langle \bar{\psi} \gamma^0
\tau^3 \psi \rangle =
f^2_\pi \mu_I (1-\lambda(\theta)^2)\\
n^-_A &=& i\langle \bar{u} \gamma^0 \gamma^5 d\rangle = - f^2_\pi
\mu_I
\lambda(\theta) \sqrt{1-\lambda^2(\theta)} \cos(\alpha)\\
n^+_A &=& \langle \bar{u} \gamma^0  d\rangle = - f^2_\pi \mu_I
\lambda(\theta) \sqrt{1-\lambda^2(\theta)} \sin(\alpha)\eeqn All
the charge densities vanish in the normal phase. In the superfluid
phase, a non-zero isospin density appears, $n_I = \frac{1}{2}
\langle\bar{\psi} \gamma^0 \tau^3 \psi\rangle$.  This is precisely
the density, which one expects to induce by applying an isospin
chemical potential $\mu_I$. At $\theta =0$ it   coincides with the
previous result \cite{Son:2000xc}. In addition, we also obtain
non-vanishing axial charge densities, $n^-_A = i\langle \bar{u}
\gamma^0 \gamma^5 d\rangle$ and $n^+_A = \langle \bar{u} \gamma^0
d\rangle$. Notice that $n^-_A$ does not vanish already at $\theta
= 0$, nevertheless, it was never discussed previously in the
literature. 
The quantity $n^-_A$ is the axial charge density,
corresponding to off-diagonal generators of the $SU(2)_A$ group,
which is both spontaneously and explicitly broken.

The   density $n^-_A$ spontaneously breaks the $U(1)_I$ symmetry
and, hence, may be considered as an order parameter alongside the
pion condensate, $\langle\pi^- \rangle = i\langle  \bar{u}
\gamma_5 d\rangle$ Note that there was no explicit chemical
potential conjugate to $n^-_A$ in the Lagrangian - once $U(1)_I$
is already spontaneously broken by $\langle \pi^- \rangle$,
$n^-_A$ is induced automatically. The reader is referred to the
paper\cite{MZ} on $N_c = N_f = 2$ $QCD$ for a few arguments, which
intuitively explain why in a system with nonvanishing $n_I$ and
$\langle \pi^-\rangle$, the second order parameter $n^-_A$
automatically appears. The quantitative behaviour of these two
order parameters is somewhat different. The pion condensate
monotonically increases with $\mu_I$ after the normal to
superfluid phase transition, and $\langle \pi^- \rangle \to
-\frac{1}{2} \pp$ for $\mu_I \gg m_\pi$. On the other hand, the
new charge density $n^-_A$ first increases after the phase
transition, reaches a peak at $\mu_I = 3^{1/4} m_\pi$, and then
decreases to $0$ for $\mu \gg m_\pi$. Of course, we always
consider only $\mu_I\ll m_\rho$.

We note that the new order parameter $n^-_A$ vanishes, in the
limit $m_q \rightarrow 0$. We expect that in the regime of
asymptotically large $\mu_I$, where analytical control is present,
and both $n_I$ and $\langle \pi^- \rangle$ are believed to be
non-vanishing, one can explicitly show that $n^-_A$ will also
appear once $m_q \neq 0$ is considered.

\section{Theta Dependence}
So far we have mostly focused on the $\mu_I$ dependence at fixed
$\theta$. In this section we would like to focus more on the
$\theta$ dependence, drawing the phase diagram in the $(\theta,
\mu_I)$ plane. We will also pay particularly careful attention to
the physics near $\theta = \pi$.

We begin by briefly reviewing the well-known $\theta$ dependence
at $\mu = 0$.  The grand canonical potential ${\Omega}(\theta)$
is, \beqn \label{EN} {\Omega}(\theta, \mu = 0) &=& - f_{\pi}^2
m^2_\pi(\theta)\eeqn 
By differentiating $\Omega(\theta)$ we can compute correlation
functions of $G \tilde{G}$, \beqn \frac{\d {\Omega}}{\d \theta}
&=& \langle i \GDG \rangle\\ - \frac{\d^2 {\Omega}}{\d \theta^2}
&=& \chi = - \int d^4x \langle T \GDG (x) \GDG
(0)\rangle_{conn}\eeqn At $\mu = 0$ we find, \beqn \nn\langle i
\GDG \rangle_{\mu = 0} &=&
-\frac{1}{4}\frac{m_u m_d}{m(\theta)} \, \sin(\theta)\,\pp\\
\chi(\mu = 0) &=& \frac{1}{4}\frac{m_u m_d}{m(\theta)}
\left(\cos(\theta) + \frac{m_u m_d}{4 m(\theta)^2}
\sin^2(\theta)\right) \pp \label{GDG0}\eeqn

Expressions (\ref{GDG0}) reflect the well-known strong $\theta$
dependence in the region $m_u \approx m_d = m_q$, $\theta \approx
\pi$. Let's introduce the asymmetry parameter, $\epsilon =
\frac{|m_u - m_d|}{m_u + m_d}$ and assume $\epsilon \ll 1$. The
$CP$ odd order parameter $\langle i G \tilde{G}\rangle$, though
apparently smooth for $\epsilon \neq 0$, experiences a steep
crossover
in the
region $|\theta - \pi| \sim \epsilon$.
Correspondingly, the topological susceptibility $\chi$ has a sharp
peak around $\theta = \pi$ of width $\Delta \theta \sim \epsilon$
and height $\chi(\pi)/|\chi(0)| = 1/\epsilon$. 


Such behaviour of the $CP$ odd order parameter $\langle i G
\tilde{G}\rangle$ strongly suggests that for $m_u = m_d$,
spontaneous breaking of $CP$ symmetry occurs at $\theta = \pi$.
This situation, known as Dashen's phenomenon, has been extensively
studied in QCD with  $N_f = 3$ and $N_f =
2$\cite{Dashen,Witten:1980sp,DiVecchia:1980ve,Creutz,Smilga,Tytgat}.
For $N_f = 3$ with $m_s \gg m_u, m_d$ it is believed that
spontaneous $CP$ breaking occurs at $\theta = \pi$ for $|m_u -
m_d| m_s < m_u m_d$.

For $N_f = 2$, the key observation\cite{Creutz,Smilga} is that
Dashen's phenomenon is not under complete theoretical control in
the effective Lagrangian (\ref{Lch}). Indeed, for a moment, we fix
$m_u = m_d$. Then, for general $\theta$, the mass term explicitly
breaks the symmetry of the effective Lagrangian (\ref{Lch}) from
$SU(2)_L \times SU(2)_R$ to $SU(2)_V$. However, for $\theta =
\pi$, the mass term in the effective Lagrangian vanishes,
restoring the symmetry to $SU(2)_L\times SU(2)_R$ and giving rise
to apparently massless goldstones: $m_\pi^2(\theta=\pi) = 0$. Yet,
no such symmetry restoration occurs in the fundamental microscopic
QCD Lagrangian at $\theta = \pi$. This contradiction is resolved
by including higher order (quadratic) mass terms in the effective
Lagrangian, which would explicitly break $SU(2)_A$ even at $\theta
= \pi$\,\cite{Smilga}. It is precisely these terms, which control
the physics of Dashen's phenomenon.

In this paper we would like to consider two different regimes. In
the first regime, one may neglect the higher order mass terms by
considering fixed $\frac{|m_u - m_d|}{m_u+m_d} \neq 0$ and
sufficiently small $m_q$. Of course, in such a regime one
automatically excludes the regions of parameter space where
Dashen's transition is realized, and may discuss only the
quantitatively steep crossover in the normal phase. The second
regime that we discuss is obtained by considering the exactly
degenerate case $m_u = m_d$. We show that this second regime
exhibits the Dashen's transition in the normal phase, which
disappears in the superfluid phase.

\subsection{Crossover Regime}

In this section we discuss the regime in which the leading order
chiral Lagrangian (\ref{Lch}) accurately describes the physics for
all $\theta$. Here we give only a brief summary of the results
concerning this regime, for further discussion see \cite{MZ}.

If the leading order (\ref{mpi}) pion mass at $\theta = \pi$,
$m_\pi^2(\theta = \pi) \propto |m_u - m_d|$, is sufficiently large
one may neglect the higher order mass terms in the effective
chiral Lagrangian. For any fixed $\frac{|m_u - m_d|}{m_u + m_d}
\neq 0$ this is achieved by considering sufficiently small $m_q$.
If the higher order mass terms are largely saturated by a third
quark of mass $m_{u,d} \ll m_s \ll \Lambda_{QCD}$, one requires,
\beq \label{meta}\frac{|m_u - m_d|}{m_u + m_d} \gg
\frac{m_{u,d}}{m_s}\sim \frac{m^2_{\pi}(\theta=0)}{M_\eta^2}\eeq
This condition is, indeed, realized in the true physical world.
If, on the other hand, the higher order terms are controlled by a
light $\eta'$ (as motivated by $N_c \rightarrow \infty$), one
needs to consider, \beq \label{mtau} \frac{|m_u - m_d|}{m_u + m_d}
\gg \frac{m \pp}{f^2_{\pi} M^2_{\eta'}} \sim
\frac{m^2_{\pi}(\theta=0)}{M^2_{\eta'}}\eeq

\begin{figure}[t]
\begin{center}
\includegraphics[angle=-90, width = 0.7\textwidth]{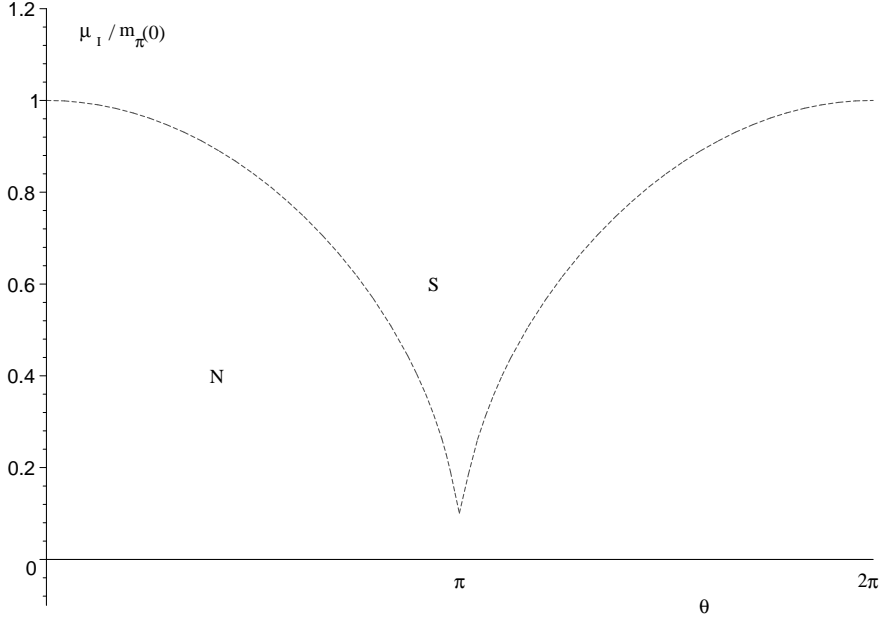}
\caption{Phase diagram of $N_f = 2$ QCD as a function of $\mu_I$
and $\theta$. Here, $\epsilon = \frac{m_u - m_d}{m_u + m_d} =
0.01$. A rapid crossover occurs in the normal phase at $\theta =
\pi$, which becomes a first order phase transition, when $m_u =
m_d$.}\label{NB}
\end{center}
\end{figure}

Let us now turn on finite $\mu_I$. Once conditions (\ref{meta}),
(\ref{mtau}) are met, all the results of previous sections hold
for any $\theta$. In particular, the transition to the superfluid
phase occurs at $\mu = m_\pi({\theta})$ (see Fig.\,\ref{NB}).

Thus, for $\mu_I < m_\pi(\theta=\pi)$ the normal phase is realized
for all $\theta$, while for $\mu_I > m_\pi(\theta=0)$ we are
entirely in the superfluid phase. Finally, if we fix $\mu_I$ with
$m_\pi(\theta=\pi) < \mu_I < m_\pi(\theta=0)$ and vary $\theta$
from $0$ to $2\pi$ we encounter two phase transitions: from normal
to superfluid phase and then back to normal. Thus, the $\theta$
dependence becomes non-analytic in this region! Since the normal
to superfluid phase transition is second order, we expect the
topological susceptibility, $\chi$ to be discontinuous across the
phase boundary. The transitions between normal and superfluid
phases occur at $\theta = \theta_c$ and $\theta = 2\pi -
\theta_c$, with the critical $\theta_c$ given by $m_\pi(\theta_c)
= \mu_I$.

In the superfluid phase, the free energy density reads, \beq
\label{EB} {\Omega}(\theta) = - \frac{1}{2} f^2_{\pi} \mu^2_I
\left(1 + \frac{m^4_\pi(\theta)}{\mu^4_I}\right).\eeq Clearly, the
$\theta$ dependence in the superfluid phase is different from that
in the normal phase (\ref{EN}). This is most clearly seen by
computing topological density and topological susceptibility
in the superfluid phase,
 \beqn \langle i \GDG \rangle &=& \frac{m_u m_d}{4
f^2_{\pi} \mu_I^2} {\pp}^2 \sin(\theta),\nn\\ \chi &=& - \frac{m_u
m_d}{4 f^2_{\pi} \mu_I^2} {\pp}^2 \cos(\theta).\label{GDGB}\eeqn
The corresponding expressions should be compared with
(\ref{GDG0}) describing the normal phase.
Focusing for a moment on $\mu_I > m_{\pi}(\theta=0)$, we see that
the $\theta$ dependence is very smooth: there is no sign of rapid
crossover in $\langle i G \tilde{G}\rangle$ near $\theta = \pi$
and the large peak in the susceptibility $\chi$ disappears.
Moreover, as $\mu_I$ increases, the $\theta$ dependence is
suppressed, as expected. This smooth $\theta $ dependence at
$\theta \sim \pi $ in the superfluid phase should be contrasted
with sharp behavior in the normal phase discussed above, see eq. (\ref{GDG0}).
 As has
been explained in detail in the parallel study on $N_c = 2$
$QCD$\cite{MZ}, the disappearance of the ``Dashen's crossover" as
$\mu_I$ increases is accomplished in the following way. First,
when $m_\pi(\theta = \pi) < \mu_I < m_\pi (\theta = 0)$, the
crossover splits into two second order normal to superfluid phase
transitions. These phase transitions replace the peak in the
topological susceptibility $\chi$ by finite jumps in $\chi$ at the
transition points: \beq \label{jump}\frac{\chi(\theta_c^+) -
\chi(\theta_c^-)}{|\chi(0)|} = \frac{m_u m_d m^2_{\pi}(0) \pp^2}{4
f^4_{\pi} \mu^6_I} \sin^2(\theta_c)\eeq Finally, once $\mu_I
> m_\pi(\theta = 0)$ no phase transitions can be triggered by
varying $\theta$, and the ``Dashen's crossover" becomes entirely
washed out.

We conclude this section by noting that we can use our results for
the topological susceptibility $\chi$ and the chiral condensate
$\langle\bar{\psi}\psi\rangle$ to study how  the Ward
Identities get saturated in different phases with arbitrary $\theta$,
\cite{DiVecchia:1980ve,Witteneta,WI,NSVZ}, \beqn
\label{WI2p}\chi = - \int d^4x \langle T \GDG(x) \,
\GDG(0)\rangle_{conn} =
\frac{1}{N^2_f}\langle \bar{\psi} M \psi\rangle + O(M^2) \\
O(M^2) = - \frac{1}{N^2_f} \int d^4x \langle T \bar{\psi} \gamma_5
M \psi(x) \, \bar{\psi} \gamma_5 M \psi(0)\rangle_{conn}.\nonumber
\eeqn This Ward Identity is related to the axial anomaly and,
thus, should not be affected by infra-red effects, such as finite
chemical potential. One can explicitly check that our results
imply that at $\theta = 0$, the identity (\ref{WI2p}) is, indeed,
straightforwardly satisfied both in the normal and superfluid
phases. However, at $\theta \neq 0$, in the superfluid phase, one
must include the $O(M^2)$ term in (\ref{WI2p}) on the same footing
as the $O(M)$ term for  the Ward Identity to be satisfied. The
reader is referred to the work \cite{MZ} where $N_c=2$ case was
discussed in detail. In the present case  with $N_c=3$ the
saturation of the Ward Identities goes precisely in the same way
as in \cite{MZ}, and therefore, we do not need to repeat it here.

\subsection{Phase Transition Regime}

In the present section, we would like to consider the degenerate
case $m_u = m_d = m$, which has not been discussed in the
companion paper\cite{MZ}. This regime is believed to support a
first order phase transition across $\theta = \pi$ at zero
chemical potential\cite{Smilga, Tytgat}. The discussion of the
crossover regime in section IVA is highly suggestive of the fact
that this phase transition disappears in the superfluid phase. We
shall now explicitly demonstrate this claim. We note that the
point $m_u = m_d$, $\theta = \pi$ might be of importance for
lattice fermions\cite{Creutz,Aoki}, as it is equivalent to a
theory where one quark mass is negative and $\theta$ parameter is
not explicitly present.

As already mentioned, in QCD with $N_f = 2$, one needs to include
second order mass terms in the effective chiral Lagrangian in
order to accurately describe physics near $m_u = m_d$, $\theta =
\pi$. As argued in \cite{Smilga}, the dominant second order mass
term is, \beq V_2(U) = -l_7 \frac{\Sigma^2}{f^4_\pi}\left(Im\Tr(M
U)\right)^2\eeq as it contains $\theta$ dependence different from
the leading mass term. Including this term in our chiral
Lagrangian, we obtain, \beq \label{Lch2} {\cal L} = \frac{1}{4}
f^2_{\pi} \Tr(\nabla_{\mu}U \nabla_{\mu}U^{\dagger}) - \Sigma
Re\Tr(M U) - l_7 \frac{\Sigma^2}{f^4_\pi} \left(Im\Tr(M
U)\right)^2\eeq

Let us review the $\theta$ dependence contained in (\ref{Lch2}) at
$\mu_I = 0$. As is known, the physics at $\mu_I = 0$ in the
neighborhood of $\theta = \pi$, crucially depends on the sign of
$l_7$. A number of arguments\cite{Smilga} suggest that $l_7$ is
positive, in particular, in the large $N_c$ limit, $l_7 \sim
\frac{f^2_\pi}{2 M^2_{\eta'}}$ \cite{Smilga,Tytgat}. We shall
assume $l_7 > 0$ for the rest of this work. In this case, the
static classical minimum of (\ref{Lch2}) is given by, $U = 1$ for
$0 \leq \theta < \pi$ and $U = -1$ for $0 < \theta \leq 2 \pi$. At
$\theta = \pi$, the classical minimum is degenerate: $U = \pm 1$,
signalling spontaneous breaking of the $P, CP$ symmetries.
Computing the value of the $CP$ order parameter, $\langle i G
\tilde{G} \rangle $, near $\theta = \pi$, \beq \langle i
\GDG\rangle_{\theta = \pm \pi} = \pm \frac{m_q}{2} |\pp|\eeq This
is exactly the result one would derive by naively setting $m_u =
m_d$ in eq. (\ref{GDG0}). Thus, once $m_u = m_d$, the rapid
crossover discussed in the previous section becomes a phase
transition.

The $\theta$ dependent mass of the three degenerate goldstones
becomes, \beq m^2_\pi(\theta) = \frac{m |\pp|}{f^2_\pi}
|\cos(\theta/2)| + \frac{2 l_7 m^2 \pp^2}{f^6_\pi}
\sin^2(\theta/2)\eeq We see that the goldstones pick up a small,
but non-vanishing, mass at $\theta = \pi$, due to the $V_2$ term
in the chiral Lagrangian.

\begin{figure}[h]
\begin{center}
\includegraphics[angle=-90, width = 0.7\textwidth]{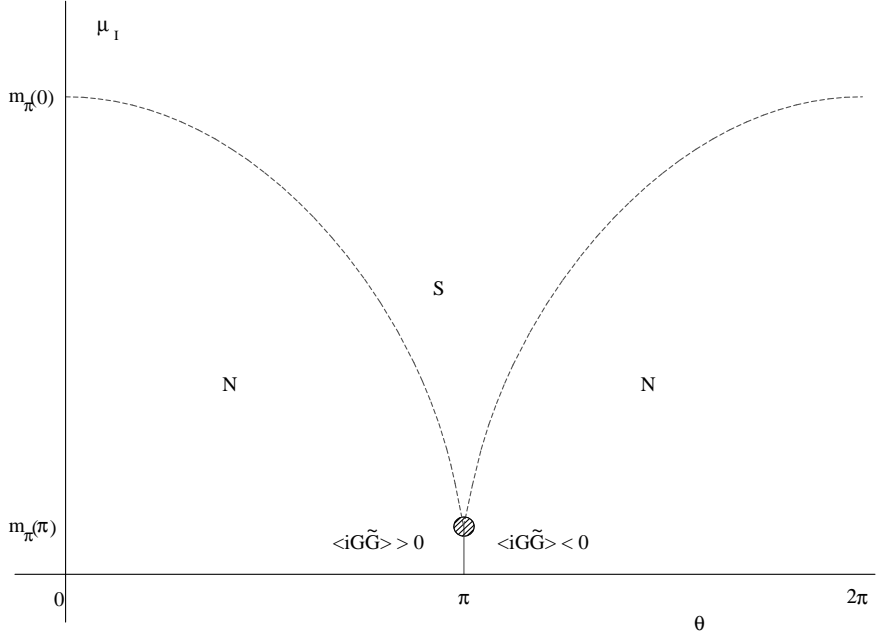}
\caption{Phase diagram of $N_f = 2$ QCD for $m_u = m_d$. Solid
line indicates a first order phase transition, while dashed lines
indicate second order phase transitions. The region near the
triple point is subject to further investigation.} \label{NBN}
\end{center}
\end{figure}

Turning on a finite chemical potential, we see that for $|\mu_I| <
m_\pi(\theta)$, we are in the normal phase, while for $|\mu_I| >
m_\pi(\theta)$, we are in the superfluid phase (see
Fig.\,\ref{NBN}).
The static
minimum $U$ of the Lagrangian (\ref{Lch2}) in both phases is again
given by expression (\ref{min}), except that now,
\beqn\lambda(\theta) = \left\{\begin{array}{cl}
sgn(\cos(\theta/2)) &\quad \textrm{normal phase}
\\
\frac{m^2_\pi(0)\cos(\theta/2)}{\mu^2_I-m^2_\pi(\pi)\sin^2(\theta/2)}
&\quad \textrm{superfluid phase}
\end{array} \right.\eeqn
The normal phase at finite $\mu_I$ again has the same physical
properties as at $\mu_I = 0$. In particular, a first order phase
transition across $\theta = \pi$ persists for $|\mu_I| <
m_\pi(\theta = \pi)$.

However, once $|\mu_I| > m_\pi(\theta = \pi)$, the Dashen's
transition splits into two second order normal to superfluid phase
transitions. The $\theta$ dependence in the superfluid phase is
very smooth. In particular, $P$ parity is not spontaneously broken
at $\theta = \pi$: one may check, \beq \langle i
\GDG\rangle_{\theta = \pi} = 0\eeq It is amusing to note, that in
the superfluid phase, spontaneous breaking of parity is shifted
from $\theta = \pi$ to $\theta = 0$.

 We observe that in the region
$m_\pi(\theta = \pi) < \mu_I < m_\pi(\theta = 0)$, the $\theta$
dependence is essentially the same as in a theory with $l_7 < 0$
and $\mu_I = 0$. Indeed, in such a theory, one would have instead
of a first order phase transition at $\theta = \pi$, two second
order phase transitions just before and after $\theta = \pi$,
accompanied by spontaneous breaking of the $SU(2)_V$
symmetry\cite{Smilga}. This is similar to the picture that we
obtain for $l_7
> 0$ and $m_\pi(\pi) <\mu_I < m_\pi(0)$, except that only the
$U(1)_I$ subgroup of $SU(2)_V$ is broken spontaneously (the other
generators of $SU(2)_V$ are explicitly broken by finite $\mu_I$).

Finally, we would like to comment regarding the triple point
$\theta = \pi$, $|\mu_I| = m_\pi(\pi)$ that appears in the phase
diagram of Fig. \ref{NBN}. At this triple point, the set of
classical minima of the Lagrangian (\ref{Lch2}) presents a sphere
$S^2$. Such degeneracy is definitely accidental, and we expect
that it will be lifted by higher order terms in the chiral
Lagrangian (most likely $O(m^3)$, $O(m \mu^2)$ terms). These
higher order terms also have the potential to change the phase
diagram in the immediate vicinity of the triple point. However, we
believe that once we are outside the window, \beq
\label{window}|\theta - \pi| < \frac{m}{\Lambda_{QCD}},
~~~~|\frac{\mu_I}{m_\pi(\pi)}-1| < \frac{m}{\Lambda_{QCD}}\eeq all
the results described above are valid. Most importantly, Dashen's
phenomenon is present in the normal phase and disappears in the
superfluid phase. Whether this disappearance occurs precisely at
the triple point (which would be the most simple scenario) or
through a more complicated series of phase transitions closely
surrounding the triple point is still an open question. In order
to answer this question one should classify all the terms in the
effective chiral Lagrangian similar to the classic
construction\cite{Leutwiler}. However, near $\theta = \pi$, the
dimensional counting rule in such a Lagrangian should be based on
the relation $m_{\pi}\sim m_q$. This is in contrast with canonical
relation $m_{\pi}^2\sim m_q$ which is the basis for the
classification scheme  presented in
\cite{Leutwiler}.\footnote{This phenomenon, when ``naively" higher
order corrections in $m_q$ start to play a crucial role has been
previously observed in eq. (\ref{WI2p}) when Ward identities have
been analyzed.} We did not attempt to analyze the corresponding
problem of classification of higher order terms in the effective
chiral Lagrangian with $\mu_I\neq 0$, $\theta\neq 0$ in the
present study. As we already  stated, outside the region
(\ref{window}) the higher order corrections can not change our
results.

To conclude the section: we have analyzed the effects of finite
$\mu_I$ on the rapid crossover, which in the absence of chemical
potential occurs at $\theta = \pi$ for $m_u \neq m_d$ and becomes
a phase transition once $m_u = m_d$. In both cases, the crossover
(first order phase transition) is replaced by two second order
phase transitions as $\mu_I$ increases. We note that if $N_f
>2$ for the  Dashen's phenomenon to happen one does not require
precise equality of the light quarks, $m_u=m_d$, rather it is
sufficient if the quark masses are close
enough\cite{Witten:1980sp,DiVecchia:1980ve}.\footnote{It is
actually possible that for $N_f = 2$ the phase transition also
occurs already for a very small, but non-zero $m_u -
m_d$\cite{Tytgat}} Our remark here is as follows: we believe that
in the case $N_f
>2$ the pattern of the replacement of the first order phase
transition by two second order phase transitions with increasing
$\mu_I$ remains the same as described in the present section.

Our last remark: Dashen's phenomenon as well as the $\theta$
dependence has been studied recently in \cite{Akemann:2001ir} in a
very different approach. The corresponding study had concentrated
on the weak coupling regime when Euclidean space time volume $L$
is small in comparison with the Goldstone mass, $L\ll
m_{\pi}^{-1}$. We emphasize that the results presented here are
valid  in the opposite regime $L\gg m_{\pi}^{-1}$ which
corresponds to the physically relevant case.

\section{Gluon Condensate}
Having determined the $\theta$ and $\mu_I$ dependence of different
condensates and densities containing
 the quark degrees of freedom,
 one can wonder if similar results  can be derived  for the gluon condensate $\<
 G_{\mu\nu}^2\>$,
 which  describes the gluon degrees of freedom. As is known, the gluon condensate represents
 the vacuum energy of the ground state in the limit $m_q=0,\,\mu=0$ and plays a crucial
 role in such models as the MIT Bag model, where a phenomenological ``bag constant" $B$
describes  the non- perturbative  vacuum energy of the system. The
question we would ideally want to answer: how will the gluon
condensate $\< G_{\mu\nu}^2\>$ (bag constant $B$) depend on $\mu,
\theta$ if the system is placed into dense matter? This question
is relevant for a number of different studies such as the equation
of state in the interior
 of neutron stars, see e.g.\cite{Alford:2004pf}, or stability of dense strangelets\cite{Zhitnitsky:2002qa}.
Of course, it is difficult to answer this question in QCD at
finite baryon density, however, the answer can be easily obtained
in QCD with $\mu_I \ll m_{\rho}$, which is
the subject of the present work. 

We work in Minkowski space in this section. We start from the
equation for the conformal anomaly, \beq \label{Tmm}
\Theta^{\mu}_{\mu} = - \frac{b g^2}{32 \pi^2} G^a_{\mu\nu} G^{a\mu
\nu} + \bar{\psi} M \psi \eeq where we have taken the standard $1$
loop expression for the $\beta$ function and $b = \frac{11}{3} N_c
- \frac{2}{3} N_f = \frac{29}{3}$, for $N_c = 3$, $N_f = 2$. As
usual, a perturbative constant is subtracted in expression
(\ref{Tmm}). 
Now, we can use the effective Lagrangian (\ref{Lch}) to calculate
the change in the trace of the energy-momentum tensor $\langle
\Theta^{\mu}_{\mu}\rangle$ due to a finite isospin chemical
potential. The energy density $\epsilon$ and pressure $p$ are
obtained from the grand canonical potential $\Omega$, \beqn
\epsilon &=& {\Omega} + \mu_I n_I, \quad p = -{\Omega} \eeqn
Therefore, the conformal anomaly implies, \beq
\label{DG2}\langle\cond\rangle_{\mu, m, \theta} -
\langle\cond\rangle_0 = -4 \left({\Omega}(\mu, m, \theta) -
{\Omega}_0\right)- \mu_I n_I(\mu, m, \theta) + \langle \bar{\psi}
M \psi \rangle_{\mu, m,\theta}\eeq Here, the subscript $0$ on an
expectation value means that it is evaluated at $\mu = m = 0, \,
\theta = 0$. The good news is that we have already calculated all
quantities on the righthand side of eq.\,(\ref{DG2}) - see
expressions (\ref{cond1}),(\ref{EN}),(\ref{EB}). Thus, in the
normal phase we obtain, \beq \label{G2N}\langle\cond\rangle_{\mu,
m,\theta} - \langle \cond\rangle_0 = - 3 m(\theta) \pp \eeq When
$\theta = 0$, (\ref{G2N}) reduces to the standard
result\cite{NSVZ}, which was derived in a different manner. As
expected, $\langle G^2_{\mu\nu}\rangle$ does not depend on $\mu$
in the normal phase. The superfluid phase is more exciting, \beq
\label{G2B} \langle\cond\rangle_{\mu, m, \theta} -
\langle\cond\rangle_0 = f_{\pi}^2 \mu^2_I \left(1 + 2
\frac{m^4_{\pi}(\theta)}{\mu^4_I}\right).\eeq It is instructive to
represent the same formula in a somewhat different way,
 \beq
\label{G2M} \langle\cond\rangle_{\mu, m, \theta} -
\langle\cond\rangle_{\mu=0, m, \theta} =  f^2_{\pi} (\mu^2_I
-m^2_{\pi}(\theta)) \left(1 - 2
\frac{m^2_{\pi}(\theta)}{\mu^2_I}\right),\eeq which makes contact
with the fact that in the normal phase, when $\mu_I\leq
m_{\pi}(\theta)$, the gluon condensate does not vary with $\mu_I$.
However, for $\mu_I\geq m_{\pi}(\theta)$, the dependence of the
gluon condensate $\langle G^2_{\mu \nu}\rangle$ on $\mu_I$ in the
superfluid phase becomes  rather interesting. The condensate
decreases with $\mu_I$ for $m_\pi < \mu_I < 2^{1/4} m_\pi$ and
increases afterwards. The qualitative difference in the behaviour
of the gluon condensate for $\mu_I \approx m_\pi$ and for $m_\pi
\ll \mu_I \ll m_{\rho}$ can be explained as follows. Right after
the normal to superfluid phase transition occurs, the isospin
density $n_I$ is small and our system can be understood as a
weakly interacting gas of pions. The pressure of such a gas is
negligible compared to the energy density, which comes mostly from
pion rest mass. Thus, $\langle\Theta^{\mu}_{\mu}\rangle$ increases
with $n_I$ and, according to the anomaly equation (\ref{Tmm}),
$\langle G^2_{\mu\nu}\rangle$ decreases. A similar decrease in
$\langle G^2_{\mu\nu}\rangle$ with baryon density is expected to
occur in ``dilute" nuclear matter (see \cite{G2orig} and review
\cite{G2rev}). On the other hand, for $\mu_I \gg m_{\pi}$, energy
density is approximately equal to pressure, and both are mostly
due to self-interactions of the pion condensate. Luckily, the
effective chiral Lagrangian (\ref{Lch}) gives us control over
these self-interactions as long as $\mu_I \ll m_{\rho}$. Such
control is largely absent in corresponding calculations of
$\langle G^2_{\mu \nu}\rangle$ in nuclear matter. As $\Delta
\epsilon \sim \Delta p$, the trace
$\langle\Theta^{\mu}_{\mu}\rangle$ decreases and the gluon
condensate increases with isospin density. Such behaviour of
$\langle G^2_{\mu \nu}\rangle$ is quite unusual, as finite quark
chemical potentials, on general grounds, are expected to suppress
the gluons.

\section{Conclusion}
The main purpose of this work was to investigate the phase diagram
of $N_f = 2$ QCD at finite $\theta$ parameter and isospin density.
We have found that the $\theta$ dependence becomes non-analytic:
for fixed $\mu_I$ of order of the pion mass, two phase transitions
of the second order occur as $\theta$ varies from $2 \pi$ to $0$.
We have also demonstrated the conjecture originally presented in
\cite{MZ}: in the limit of degenerate quark masses, spontaneous
$P$ breaking occurs in the normal phase, but is absent in the
superfluid phase. For $m_u = m_d$, a first order  transition
across $\theta = \pi$ is present in the normal phase, but
disappears in the superfluid phase by splitting into two second
order normal to superfluid transitions. The precise details of the
neighborhood of the triple point where such splitting takes place
remain to be determined.

There are a few more interesting observations which deserve to be mentioned here:\\
  a) Knowledge of $\theta$ dependence of different condensates allows one to calculate the topological susceptibility
  and other  interesting correlation functions as a function of $\mu$. Corresponding  Ward Identities at nonzero $\mu_I$ are satisfied in a quite nontrivial way, and
can be tested on the lattice.\\
b)  Physics of gluon degrees of freedom and $\mu_I$ dependence of
the gluon condensate can also be tested on the lattice. The
behavior of the gluon condensate as a function of $\mu_I$ is very
nontrivial, as has been explained in the text. Nevertheless, our
prediction is robust in a sense that it is based exclusively on
the chiral dynamics and no additional assumptions have been made
to derive the corresponding expression. Our formulae might be
useful for the lattice simulations when one tries to extrapolate the
results to the chiral limit at nonzero $\mu_I$.

Finally, we should emphasize  that all results presented above are
valid only for very small chemical potentials $ \mu_I\ll
\Lambda_{QCD}$ when the chiral effective theory is justified. For
larger chemical potentials  we expect a transition to a deconfined
phase at $ \mu_I \simeq 5 \Lambda_{QCD}$ \cite{TZ}.

\section*{Acknowledgements}
We would like to thank M. Stephanov for useful discussions. This
work
 was supported, in part, by the Natural Sciences and Engineering
Research Council of Canada.


\begin{thebibliography}{X}
\bibitem{Dashen} R.~Dashen, Phys.\ Rev.\ D{\bf 3}{(1971) }, {1879}
\bibitem{Witten:1980sp}
E.~Witten, \emph{Large {$N$} chiral dynamics}, Ann. of Phys. {\bf
128} (1980), {363}.

\bibitem{DiVecchia:1980ve}
P.~Di~Vecchia and G.~Veneziano,
Nucl. Phys. {\bf B 171}(1980),{253}.

\bibitem{Creutz} M. Creutz, Phys.\ Rev.\ D{\bf 52}{(1995) }, {2951}; Phys. Rev. Lett. {\bf 92} (2004) 162003;
Phys. Rev. Lett. {\bf 92} (2004) 201601.
\bibitem{Smilga}
  A.~V.~Smilga,
  Phys.\ Rev.\ D {\bf 59}, 114021 (1999).
\bibitem{Tytgat}
M.~H.~G.~Tytgat,
Phys.\ Rev.\ D {\bf 61}, 114009 (2000) [arXiv:hep-ph/9909532].

\bibitem{ARF}
M.~Alford, K.~Rajagopal, and F.~Wilczek,
Phys.\ Lett.\  {\bf B 422} (1998) 247;\\ 
R.~Rapp, T.~Sch\"afer, E.~V.~Shuryak, and M.~Velkovsky,
Phys.\ Rev.\ Lett.\  {\bf 81} (1998) 53.

\bibitem{kogut1} J.B. Kogut, M.A. Stephanov, and D. Toublan, Phys. Lett. {\bf
    B 464} (1999) 183-191.

\bibitem{kogut2} J.B. Kogut, M.A. Stephanov, D. Toublan, J.J.M. Verbaarschot,
  and A. Zhitnitsky, Nucl. Phys. {\bf B 582} (2000) 477-513.

\bibitem{Splittorff:2000mm}
K.~Splittorff, D.~T.~Son and M.~A.~Stephanov, ``QCD-like theories
at finite baryon and isospin density,'' Phys.\ Rev.\ D {\bf 64},
016003 (2001). 

\bibitem{Son:2000xc}
D.~T.~Son and M.~A.~Stephanov,
Phys.\ Rev.\ Lett.\  {\bf 86}, 592 (2001).

\bibitem{Son:2000by}
D.~T.~Son and M.~A.~Stephanov,
Phys.\ Atom.\ Nucl.\  {\bf 64}, 834 (2001) [Yad.\ Fiz.\  {\bf 64},
899 (2001)].

\bibitem{MZ}
M.~A.~Metlitski and A.~R.~Zhitnitsky,
[arXiv:hep-ph/0508004] (accepted to Nucl.\ Phys.\ {\bf B})

\bibitem{Witteneta}
E. Witten, Nucl. Phys. {\bf B156} (1979) 269.



\bibitem{WI}
R.Crewther, Phys. Lett. {\bf B70} (1977)  349.
\bibitem{NSVZ} V. Novikov,  M.A. Shifman, A.I. Vainshtein, and V.I. Zakharov, Nucl. Phys.
{\bf B166} (1980) 493.
\bibitem{Aoki} S. Aoki, Phys. Lett. {\bf B190}
(1987)  140.
\bibitem{Leutwiler} J. Gasser and H. Leutwyler, Ann. Phys. {\bf 158},  142 (1984).
\bibitem{Akemann:2001ir}
  G.~Akemann, J.~T.~Lenaghan and K.~Splittorff,
  Phys.\ Rev.\ D {\bf 65}, 085015 (2002)
  [arXiv:hep-th/0110157].
\bibitem{Alford:2004pf}
  M.~Alford, M.~Braby, M.~W.~Paris and S.~Reddy,
  [arXiv:nucl-th/0411016].
\bibitem{Zhitnitsky:2002qa}
  A.~R.~Zhitnitsky,
  JCAP {\bf 0310} (2003) 010.
  [arXiv:hep-ph/0202161].
\bibitem{G2orig}
T.~D.~Cohen, R.~J.~ Furnstahl, and D.~K.~ Griegel, Phys.\ Rev.\
{\bf C45}, 1881 (1992).
\bibitem{G2rev}
E.~G.~Drukarev, M.~G.~Ryskin and V.~A.~Sadovnikova, Prog.\ Part.\
Nucl.\ Phys.\ {\bf 47}, 73 (2001).


\bibitem{TZ}
  D.~Toublan and A.~R.~Zhitnitsky,
  [arXiv:hep-ph/0503256].















\end{thebibliography}
\end{document}